# Observation of the out-of-plane orbital antidamping-like torque

Zeyang Gong[1,†], Fu Liu[1,†], Xinhong Guo[1], Changjun Jiang[1]*

[1] Key Laboratory for Magnetism and Magnetic Materials, Ministry of Education, Lanzhou University, Lanzhou 730000, China

[†]These authors contributed equally to this work.

Corresponding author. E-mail address: *jiangchj@lzu.edu.cn

## Abstract

The out-of-plane antidamping-like orbital torque fosters great hope for high-efficiency spintronic devices. Here we report experimentally the observation of out-of-plane antidamping-like torque that could be generated by *z*-polarized orbital current in ferromagnetic-metal/oxidized Cu bilayers, which is presented unambiguously by the magnetic field angle dependence of spin-torque ferromagnetic resonance signal. The oxidized Cu thickness dependence of orbital torque ratios highlights the interfacial effect would be responsible for the generation of orbital current. Besides that, the oxidized Cu thickness dependence of damping parameter further proves the observation of antidamping-like torque. This result contributes to enriching the orbital-related theory of the generation mechanism of the orbital torque.

# 1 Introduction

Orbital current, a flow of orbital angular momentum (OAM), have attracted wide interest in the recent past.[1-4] Orbital current is defined as the directional flow of carriers carrying OAM, which carriers angular momenta polarized in different directions according to the different direction of their own OAM.[5] Generally, there are two mechanisms giving rise to the generation of orbital current, orbital Hall effect (OHE)[6-7] and orbital Rashba effect (ORE).[8-9] OHE refers to a longitudinal electric field inducing an excitation of orbital current in a transverse direction. While the ORE allows the generation of chiral OAM texture in **k** space via OAM dependence of energy splitting analogous to spin Rashba effect (SRE). Notably, unlike SRE, the ORE can still exist without spin-orbit coupling (SOC), with the broken inversion symmetry alone adequate to its emergence.[10] Generally, the resulting orbital current is polarized along y direction when the charge current along x direction, while the orbital current flows toward z direction similar to spin current.[11] This orbital current (to distinguish it from spin current, denoting as $o_y$) can generate an in-plane antidamping-like torque (orbital torque, OT) $\tau_{DL}$ that contributes to highly efficient magnetization switching of thin films similar to spin torque.[12] Besides, the out-of-plane antidamping-like torque possess potential applications, in which the generation mechanism of this torque is worthing studying for the development of orbital engineering.

Recently, the out-of-plane antidamping-like torque is widely reported in spin-related systems that generated by $\sigma_z$-polarized spin current, which prevails in specific system with lower crystalline or magnetic symmetry, such as anti-ferromagnetic (AFM) materials $Mn_3GaN$[13], two-dimensional semimetals $WTe_2$[14] as well as ferromagnet/ferroelectric multiferroic (FM/FE) heterostructures interface.[15] Meanwhile, the confirmation of orbital current becomes more popular issue, where more and more systems are confirmed that exist orbital torque, such as Ta/Ni[1], Ru/Ni[16], Nb/Co[17], etc. In these systems, the orbital current induced by OHE and the OT is proved to exist in the form of in-plane damping-like torque, which is generated by $o_y$. Besides, the orbital current has been theoretically and experimentally verified in naturally oxidized Cu

($CuO_x$) film induced by ORE[18-19], which the form of orbital current is still $o_y$. However, the out-of-plane polarized orbital current ($o_z$) is rarely reported despite its rich orbital physics in orbitronics.

Herein, the $o_z$-polarized orbital current and out-of-plane antidamping-like torque are observed in FM/$CuO_x$ bilayers via spin-torque ferromagnetic resonance (ST-FMR). The orbital current generated by ORE in the FM/$CuO_x$ interface as shown in Fig. 1a, the induced orbital current are injected into FM and next converted to spin through SOC of FM. From the magnetic field dependence of ST-FMR spectra, excepting for $o_y$-polarized orbital current, we further observe out-of-plane antidamping-like torque generated by $o_z$-polarized orbital current. After fitting the ST-FMR antisymmetric resonance component $V_a$ vs in-plane magnetic-field angle $\varphi$ and symmetric component $V_s$ vs $\varphi$, respectively, we conclude the magnitude of various OT and the equivalent OT efficiency. Besides that, we also characterize the FM species dependence of OT efficiency, which result coincides with orbital-related theory.

## 2. Experimental

The FM/Cu thin films were grown on thermally oxidized Si substrates via magnetron sputtering at room temperature. For clarifying the physical mechanism of the OT, two different types of bilayers were prepared: Co(8 nm)/Cu($t_{Cu}$) and $Fe_{20}Ni_{80}$(8 nm)/Cu($t_{Cu}$) ($Fe_{20}Ni_{80}$ = Permalloy = Py). The devices were prepared by exploiting shadow mask prior to the deposition and the dimension is 100 × 600 μm² rectangular strips. The FM/$CuO_x$ was obtained via the FM/Cu bilayers exposed to the laboratory ambient.

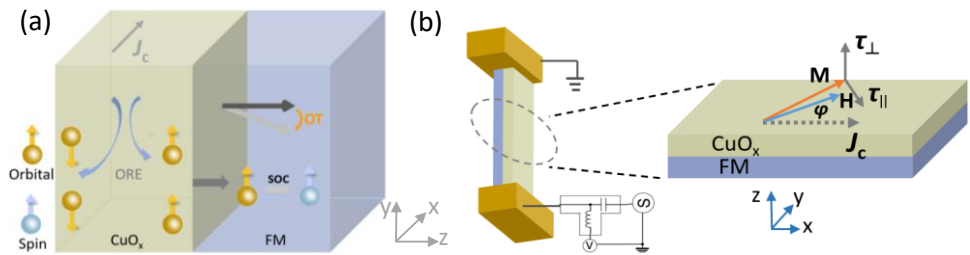

**Fig. 1** (a) Schematic illustration for the mechanism of orbital torque in FM/$CuO_x$. (b) Schematic of the device geometry and the circuit for ST-FMR measurements.

To measure the charge-to-orbital conversion, we use the ST-FMR technique as shown in Fig. 1b. In the ST-FMR measurement, a radio frequency (RF) microwave current was applied in the FM/CuO$_x$ interface producing alternating torques into the FM layer, which excites the FM magnetization to precess and drives the resistance oscillations owing to the anisotropic magnetoresistance (AMR) of the FM layer when the microwave frequency and the external magnetic field satisfy the ferromagnetic resonance (FMR) condition[20],

$$f = \frac{\gamma}{2\pi}\sqrt{H(H + 4\pi M_{eff})}. \tag{1}$$

Here, $M_{eff}$ is the effective saturation magnetization and $\gamma$ is the gyromagnetic ratio. We measure a direct current (DC) voltage $V_{dc}$ across the stripe that couples the oscillating resistance and the RF current. In the ST-FMR measurements, this DC signal $V_{dc}$ can be fitted with the equation[21]:

$$V_{dc} = V_a \frac{\Delta H(H-H_r)}{(H-H_r)^2+\Delta H^2} + V_s \frac{\Delta H^2}{(H-H_r)^2+\Delta H^2}, \tag{2}$$

where $\Delta H$ is the resonance linewidth, and $H_r$ is the resonance field. $V_s$ and $V_a$ are the magnitude of the symmetric and antisymmetric component, respectively. The out-of-plane ($\tau_\perp$) and in-plane ($\tau_\parallel$) torques defined in Fig. 1b could be obtained individually, as the antisymmetric ST-FMR resonance component $V_a$ and symmetric component $V_s$ are proportional to the amplitude of the out-of-plane and in-plane torques. These components relate to the two torques as follows[13]:

$$V_s = -\frac{I_{RF}}{2}\left(\frac{dR}{d\varphi}\right)\frac{1}{\alpha\gamma(2H_r+\mu_0 M_{eff})}\tau_\parallel, \tag{3}$$

$$V_a = -\frac{I_{RF}}{2}\left(\frac{dR}{d\varphi}\right)\frac{\sqrt{1+\mu_0 M_{eff}/H_r}}{\alpha\gamma(2H_r+\mu_0 M_{eff})}\tau_\perp. \tag{4}$$

where $\varphi$ is the in-plane magnetic field angle defined in Fig. 1b. d$R$/d$\varphi$ is due to the AMR in the FM. $I_{RF}$ is the radio microwave current. $\mu_0$ is the permeability in vacuum, $\alpha$ is the Gilbert damping coefficient.

## 2 Results and discussion

Similar to spin-related systems, considering only orbital Hall effect (or the orbital Rashba effect and Oersted field), the out-of-plane and in-plane torques could be generally given $\boldsymbol{m}\times\boldsymbol{o_y}$ and $\boldsymbol{m}\times(\boldsymbol{o_y}\times\boldsymbol{m})$, respectively.[22] Under the circumstances, analogous to systems with 2-fold rotational symmetry, if $\boldsymbol{m}$ is reversal by applying a negative magnetic field equivalent to rotating the in-plane magnetic field angle $\varphi$ (in regard to x direction) by 180°, $V_{dc}$ must satisfy same amplitude but opposite sign, that is $V_{dc}(H) = -V_{dc}(-H)$. Any difference in resonance line shape between $V_{dc}(H)$ and $-V_{dc}(-H)$ represents the existence of an additional torque. Generally, besides $o_y$-induced OT, the $o_z$-induced OT, with features of more comparable $V_s$ signals (both the magnitude and sign are taken into account) and distinct $V_a$ signals, which may be considered.

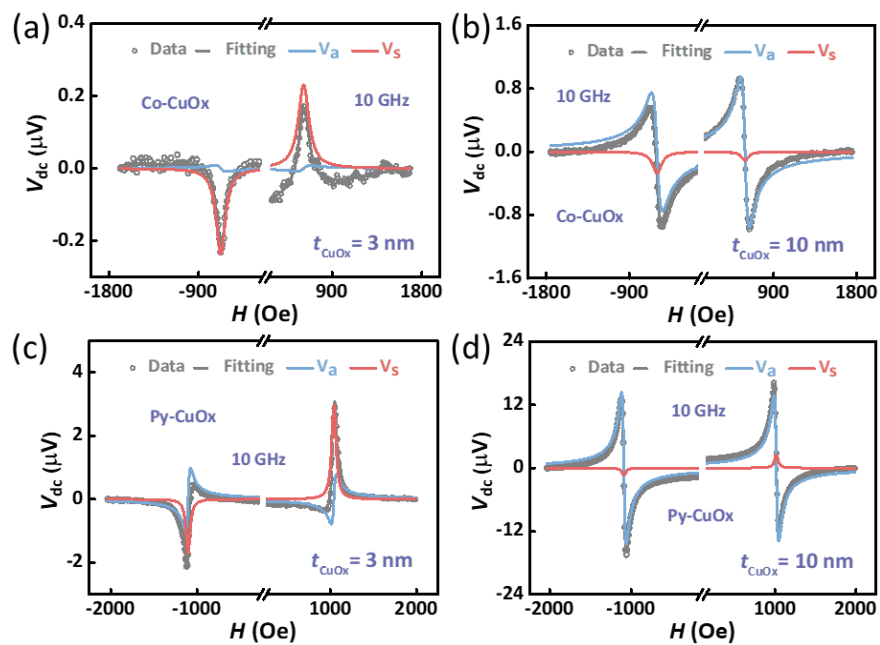

**Fig. 2** ST-FMR resonance signal for (a) Co/CuO$_x$(3 nm), (b) Co/CuO$_x$(10 nm), (c) Py/CuO$_x$(3 nm) and (d) Py/CuO$_x$(10 nm).

Fig. 2a shows the representative ST-FMR resonance signal $V_{dc}$ of the Co/CuO$_x$(3 nm) at frequency of 10 GHz with the current flow for the magnetic field angle $\varphi = 45°$ (sweeping from the positive and negative magnetic field) at room temperature. According to previous analysis, comparing the ST-FMR resonance signals under positive and negative magnetic field, $V_a$ signals with opposite signs reflects an existence of $o_z$-induced OT. Moreover, as presented in Fig. 2b for Co/CuO$_x$(10 nm), $V_s$ signals with same signs but smaller magnitude relative to $V_a$ means weaker contribution from $o_z$-induced OT when the thickness of CuO$_x$ reaching to 10 nm. One representative characteristic of the magnitude of OT is strongly dependent on material variation of a FM.[2] Thus, we perform similar ST-FMR measurements for Py/CuO$_x$(3 nm) bilayers as shown in Fig. 2c, distinct $V_s$ and $V_a$ signals in magnitude provides obvious evidence for the existence of $o_z$-induced OT. While as for Py/CuO$_x$(10 nm) bilayers, there is no obvious difference between $V_{dc}(H)$ and $-V_{dc}(-H)$, which indicates the weak $o_z$-induced OT.

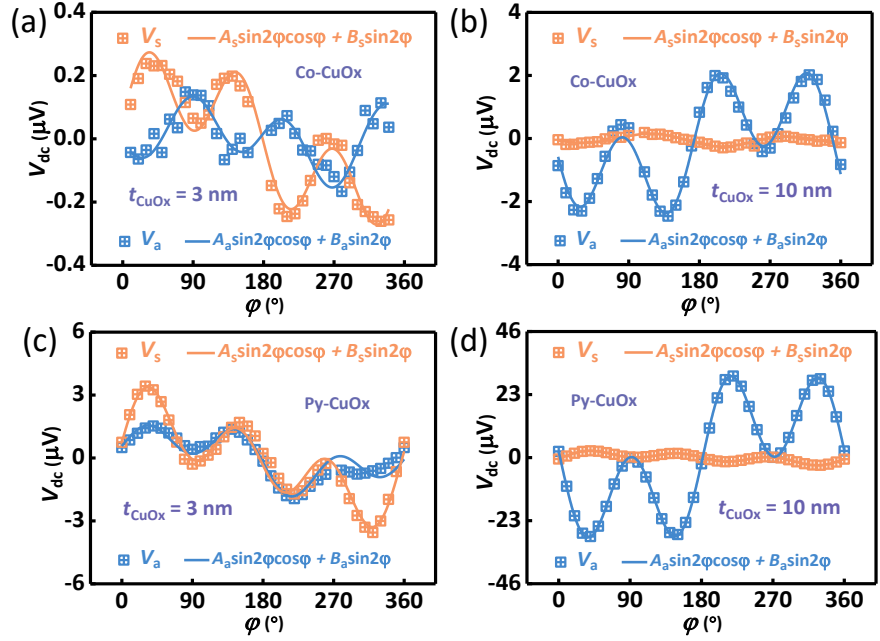

**Fig. 3** In-plane magnetic field angle $\varphi$ dependence of $V_s$ and $V_a$ for (a) Co/CuO$_x$(3 nm), (b) Co/CuO$_x$(10 nm), (c) Py/CuO$_x$(3 nm) and (d) Py/CuO$_x$(10 nm).

To analyse the torque components and magnitude in FM/CuO$_x$ systems, we consider the full angular dependence of the ST-FMR signal as an external magnetic field. The ST-FMR measurement is conducted for values of the in-plane magnetic-field angle $\varphi$ from 0º to 360º at an excitation frequency of 10 GHz. To obtain $V_s$ and $V_a$, the resonance spectra are fitted via Eq. (2). The angular dependence of $V_s$ and $V_a$ are exhibited in Fig. 3a for Co/CuO$_x$(3 nm) bilayers, which can be fitted by the variation of Eq. (3) and (4) that is $V_s \propto \sin(2\varphi)\tau_\parallel$ and $V_a \propto \sin(2\varphi)\tau_\perp$. Traditionally, analogous to spin-related systems, the current-induced orbital torques follow a cos ($\varphi$) behavior due to the orbital Hall effect and/or Oersted field[20,23], which leads to a total angular dependence of the form sin (2$\varphi$) cos ($\varphi$) for both $V_s$ and $V_a$. However, it is notable that the angular dependence both $V_a$ and $V_s$ for the Co/CuO$_x$(3 nm) bilayers need to be well fitted through adding extra term, that is:

$$V_s = A_s \sin(2\varphi)\cos(\varphi) + B_s \sin(2\varphi) \tag{5}$$

$$V_a = A_a \sin(2\varphi)\cos(\varphi) + B_a \sin(2\varphi) \tag{6}$$

where $A_{s(a)}$ and $B_{s(a)}$ represent the magnitude of OT $\tau_{y,AD(FL)}$ and $\tau_{z,FL(AD)}$ generated by orbital current that are polarized along y direction ($\boldsymbol{o_y}$) and z direction ($\boldsymbol{o_z}$), respectively. It is essential to point that the out-of-plane field-like torque $\tau_{y,FL}$ is deemed to the contribution from current-induced Oersted field.[13] In other words, the magnitude of corresponding OT could be obtained by fitting $V_{a(s)}$ as functions of $\varphi$. As shown in Fig. 3a, better fit through two terms further indicates the existence of $\boldsymbol{o_z}$-induced OT. However, as for Co/CuO$_x$(10 nm) bilayers in Fig. 3b, smaller $V_s$ and a trend of close to $\sin(2\varphi)\cos(\varphi)$ for $V_a$ means weaker $\boldsymbol{o_z}$-induced OT. Moreover, in Py/CuO$_x$ bilayers, similar behavior has been observed, that is, strong $\boldsymbol{o_z}$-induced OT appears at a CuO$_x$ thickness ($t_{CuOx}$) of 3 nm in Fig. 3c, while weaken at 10 nm in Fig. 3d.

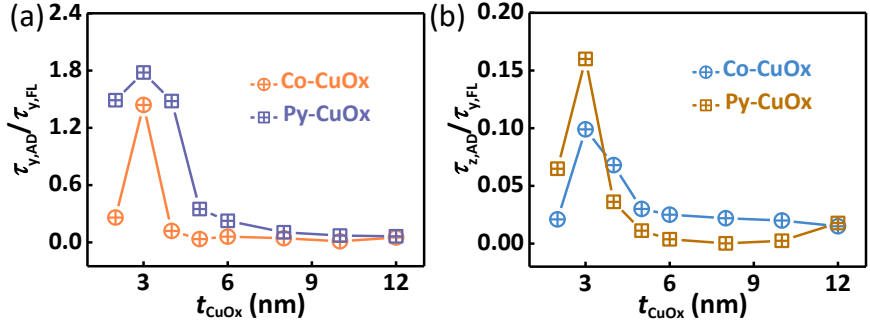

**Fig. 4** (a) $\frac{\tau_{y,AD}}{\tau_{y,FL}}$ and (b) $\frac{\tau_{z,AD}}{\tau_{y,FL}}$ corresponding to in-plane antidamping-like torque and out-of-plane antidamping-like torque generated by orbital polarization along y and z direction for Co/CuO$_x$ and Py/CuO$_x$ bilayers, respectively.

Furthermore, to obtain the proportion of antidamping-like torque in the generated torques, we employ the OT ratios $\frac{\tau_{y,AD}}{\tau_{y,FL}}$ and $\frac{\tau_{z,AD}}{\tau_{y,FL}}$ corresponding to in-plane antidamping-like torque and out-of-plane antidamping-like torque generated by orbital polarization along $y$ and $z$ direction, respectively. We extract the magnitude of OT from $V_{a(s)}$ vs $\varphi$ under different $t_{CuOx}$ for Co/CuO$_x$ and Py/CuO$_x$ bilayers and conclude the OT ratios, showing $t_{CuOx}$ dependence of $\frac{\tau_{y,AD}}{\tau_{y,FL}}$ in Fig. 4a. when the $t_{CuOx}$ from 2 nm to 3 nm, the $\boldsymbol{o_y}$-induced OT ratios $\frac{\tau_{y,AD}}{\tau_{y,FL}}$ increases dramatically and peaks when $t_{CuOx}$ reaches 3 nm, highlighting that the orbital current stems from FM/CuO$_x$ interface. For thicker samples ($t_{CuOx} > 4$ nm), the system becomes FM/Cu/CuO$_x$, it exhibits a monotonic decrease of OT ratios as $t_{CuOx}$ increases, which can be attributed to either the current shunting of low resistivity Cu or the blocking of the orbital current in non-oxidized Cu with filled 3d-shell.[18] Moreover, the $o_z$-induced OT ratios ($\frac{\tau_{z,AD}}{\tau_{y,FL}}$) also shows similar trend as shown in Fig. 4b. It is obvious that there exists out-of-plane antidamping-like torque when $t_{CuOx}$ is 3 nm. We contribute the generation of orbital current to ORE[19] and spin-vorticity coupling (SVC).[24] Note that the oxygen gradient has a vital effect on OT and the modulation of orbital current as the electronic configuration of $CuO_x$ alters to $d^8$ or $d^9$ not $d^{10}$. According to above result, we conclude that thinner Cu layer ($t_{Cu} < 4$ nm) is all oxidized to $CuO_x$ and adjacent to FM layer, where the OAM is absorbed by the magnetization of FM layer. However, the thicker Cu layer, where the system exhibits the structure of FM/Cu-$CuO_x$ due to shorter length scale of the oxygen gradient (~ 3-5 nm), hence the orbital current cannot efficiently diffuse through the non-oxidized Cu due to intrinsic quenching of the OAM, which could be ascribed to the negligible $d$ character of non-oxidized Cu near the Fermi level.[19]

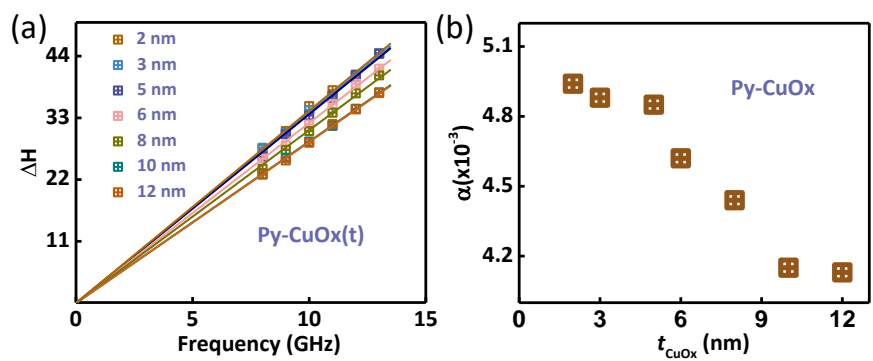

**Fig. 5** (a) Frequency dependence of linewidth $\Delta H$ for different $CuO_x$ thicknesses. (b) $CuO_x$ thickness dependence of damping parameter $\alpha$.

To further identify the existence the antidamping-like torque in the system, we carry out microwave frequency $f$ dependence of ST-FMR measurement for Py/$CuO_x$($t_{CuOx}$) bilayers to obtain the variety of damping.[25] Following, we extract resonance linewidth $\Delta H$ under different frequency for each sample, showing the $f$ dependence of $\Delta H$ in Fig. 5a and fitted by following equation[26]: $\Delta H = \Delta H_0 + 4\pi\alpha f/\gamma$, where $\Delta H_0$ and $\alpha$ can be quantitatively obtained by linear fitting. Almost-zero inhomogeneous linewidth broadening $\Delta H_0$ implies small roughness of each sample.

Moreover, when the thickness of $CuO_x$ gradually increase, the slope of $\Delta H$ vs $f$ decrease, which means a similar trend of $\alpha$ and further characterizes in Fig. 5b. With the increasing $CuO_x$ thickness, we find a monotonic decrease of damping $\alpha$, which proves the existence of anti-damping[25] (decrease of $\alpha$) and attributes to interfacial OT by ORE at $Py/CuO_x$ system. We note that similar trend was observed in Py/Ag/Bi[26], Py/Cu/Ta[27] and Ta/Py[25] interfacial Rashba systems, where the decrease of $\alpha$ indicates the generation of antidamping-like torque.

## 3 Conclusions

In conclusion, the generation of out-of-plane antidamping-like torque is investigated based on orbital Rashba effect in $FM/CuO_x$ bilayers by employing spin-torque ferromagnetic resonance. The magnetic field dependence of resonance signal proves the existence of both **y**- and **z**-polarized orbital current. The $o_y$- and $o_z$- induced orbital torque ratios $\frac{\tau_{y,AD}}{\tau_{y,FL}}$ and $\frac{\tau_{z,AD}}{\tau_{y,FL}}$ all increase dramatically and peak when the thickness of $CuO_x$ reaches 3 nm and then monotonically decrease for thicker $CuO_x$, highlighting that the orbital current originates in $FM/CuO_x$ interface and strong antidamping-like torque present in 3-nm $CuO_x$ systems, which further be proved by $CuO_x$ thickness dependence of damping parameter $\alpha$. This result provides a physic understanding of the generation out-of-plane antidamping-like torque by orbital current and opens a route to generating efficient OT by orbital engineering.

## Conflict of Interest

The authors declare no conflict of interest.

## Acknowledgements

This work was supported by the National Natural Science Foundation of China (Grant No. 52271179), the Natural Science Foundation of Gansu Province (Grant No. 21JR7RA472).